\documentclass[aps,prd,floatfix,superscriptaddress,preprintnumbers]{revtex4}
\usepackage{amssymb}
\usepackage{amsmath}
\usepackage{amsfonts}
\usepackage{epsfig}             
\usepackage{graphicx}
\usepackage{tabularx}
\usepackage{color}
\newcommand{\eq}{\begin{eqnarray}}
\newcommand{\en}{\end{eqnarray}}

\newcommand{\ra}{\rangle}

\newcommand{\bfk}{{\bf k}_{\perp}}

\newcommand{\bfl}{{\boldsymbol\ell}_{\perp}}

\begin{document}

\preprint{SLAC-PUB-17676}

\title{Predictions for the Sivers Single-Spin Asymmetry from Holographic QCD}

\author{Valery E. Lyubovitskij}
\affiliation{Institut f\"ur Theoretische Physik,
Universit\"at T\"ubingen,
Kepler Center for Astro and Particle Physics,
Auf der Morgenstelle 14, D-72076 T\"ubingen, Germany}
\affiliation{Departamento de F\'\i sica y Centro Cient\'\i fico
Tecnol\'ogico de Valpara\'\i so-CCTVal, Universidad T\'ecnica
Federico Santa Mar\'\i a, Casilla 110-V, Valpara\'\i so, Chile}
\affiliation{Millennium Institute for Subatomic Physics at
the High-Energy Frontier (SAPHIR) of ANID, \\
Fern\'andez Concha 700, Santiago, Chile}
\author{Ivan Schmidt}
\affiliation{Departamento de F\'\i sica y Centro Cient\'\i fico
Tecnol\'ogico de Valpara\'\i so-CCTVal, Universidad T\'ecnica
Federico Santa Mar\'\i a, Casilla 110-V, Valpara\'\i so, Chile}
\author{Stanley J. Brodsky} 
\affiliation{SLAC National Accelerator Laboratory, 
Stanford University, Stanford, CA 94309, USA}

\begin{abstract}

A new approach to nonperturbative QCD, {\it holographic light-front QCD}, provides 
a comprehensive model for hadron dynamics and spectroscopy, incorporating color confinement,
a universal hadron mass scale, the prediction of a massless pion in the chiral limit, and connections 
between the spectroscopy of mesons, baryons and tetraquarks across the full hadron spectrum. 
In this article we present predictions for the Sivers asymmetry and related transverse momentum 
distributions for the proton based on the light-front wavefunctions of the baryon eigenstate 
predicted by holographic QCD.

\end{abstract}

\maketitle 

 A novel approach to nonperturbative QCD dynamics, 
{\it holographic light-front QCD}~\cite{Brodsky:2014yha,Brodsky:2007hb}, has led 
to effective semiclassical relativistic bound-state equations for hadrons of arbitrary spin; 
it incorporates fundamental hadronic properties which are not apparent from the QCD Lagrangian, 
including color confinement, the emergence of a universal hadron mass scale, the prediction of 
a massless pion in the chiral limit, and remarkable connections between the spectroscopy of 
mesons, baryons and tetraquarks across the full hadronic spectrum. The light-front (LF) holographic formalism 
is based on the combination of superconformal quantum mechanics~\cite{deAlfaro:1976vlx,Fubini:1984hf},  
light-front quantization~\cite{Dirac:1949cp}, and the holographic embedding on a higher dimensional 
gravity theory~\cite{Maldacena:1997re} (gauge/gravity correspondence), leading to new analytic insights
into the structure of hadrons and their dynamics.

Holographic QCD also provides analytic predictions for the boost invariant
light-front wave functions (LFWFs) $\psi_H(x_i, k_{\perp i}, \lambda_i )$ of each 
hadronic eigenstate, the solutions of the light-front Schr\"odinger equations.
The knowledge of the LFWFs allows one to predict dynamical properties of the hadrons 
such as form factors, the distribution amplitudes underlying exclusive hadronic amplitudes, 
the structure functions underlying deep inelastic processes, 
etc.~\cite{Brodsky:2014yha,Brodsky:2007hb,Sufian:2016hwn,deTeramond:2018ecg,%
Gutsche:2016gcd,Gutsche:2013zia,Lyubovitskij:2020xqj,Lyubovitskij:2020otz}.                              

\begin{figure}[hb]
\begin{center}
\vspace*{4cm}
\epsfig{figure=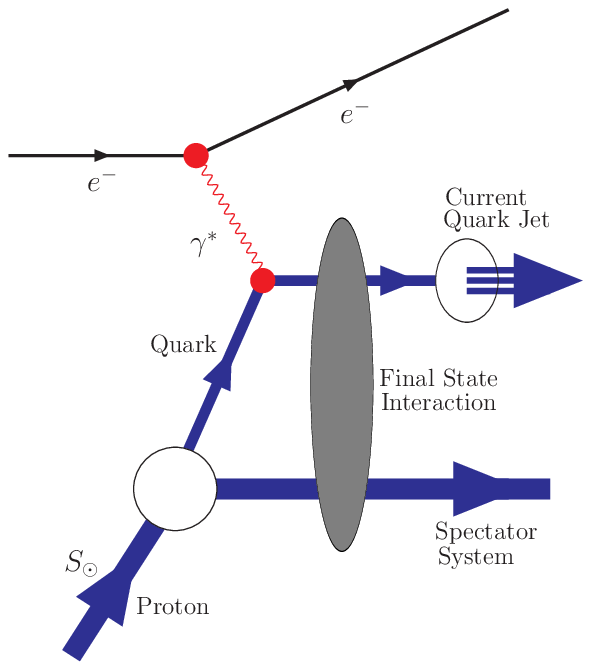,scale=.6}
\caption{Final-state interaction in the lepton SIDIS process 
$\ell p^\uparrow \to \ell' H X$. 
\label{fig1}}
\end{center}

\begin{center}
\vspace*{4cm}
\epsfig{figure=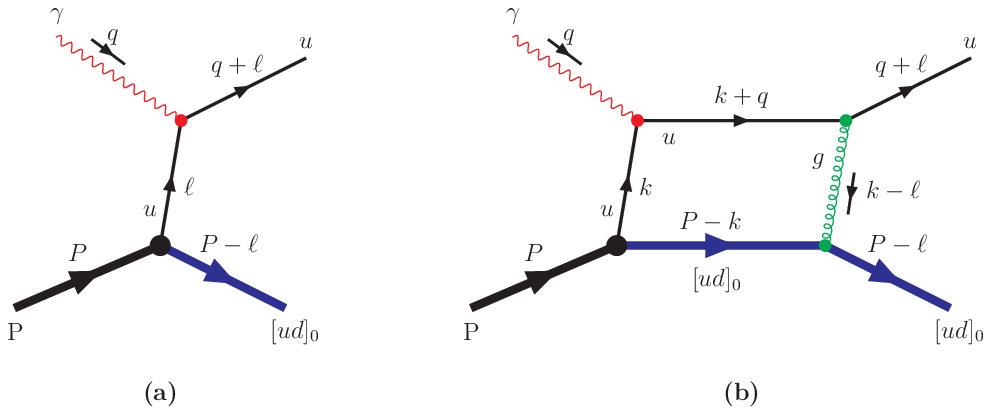,scale=.6}
\caption{Tree (a) and one-loop (b) diagrams contributing to the 
electroproduction process $\gamma^* p \to u[ud]_0$. 
\label{fig2}}
\end{center}
\end{figure}

In this paper we will use the LF Holographic  QCD approach
to predict the Sivers asymmetry as formulated in Ref.~\cite{Brodsky:2002cx}. 
In particular, it was shown~\cite{Brodsky:2002cx} that a single-spin asymmetry (SSA)  
in the semi-inclusive deep inelastic (SIDIS) electroproduction $\gamma^* p \to H X$ 
arises from final-state interactions. 
We will focus on the subprocess of the lepton SIDIS process $\ell p^\uparrow \to \ell' H X$ 
shown in Fig.~\ref{fig1}, where the quark jet produces a leading pion,  
including the final-state interactions arising from gluonic exchange.
In LF Holography, the valence state of the proton is the bound 
state of struck $u$ quark and a scalar spectator diquark $[ud]_0$ --- 
the quark-diquark model of nucleon.
The interference of tree and one-loop graphs (see Fig.~\ref{fig2}) contributing to 
the electroproduction reaction $\gamma^* p \to u[ud]_0 $ generates 
the single-spin asymmetry (SSA) of the proton~\cite{Brodsky:2002cx}.

The first experimental measurement of the Sivers effect was performed by the HERMES Collaboration 
at DESY in the study of the electroproduction of pion in the SIDIS on the transversally 
polarized hydrogen gas target~\cite{HERMES:2004mhh} and was later confirmed in 
Ref.~\cite{HERMES:2009lmz}. Then, in Refs.~\cite{COMPASS:2005csq,COMPASS:2010hbb}  
the COMPASS Collaboration at SPS CERN reported results on the Collins
and Sivers asymmetries for charged hadrons produced in SIDIS
of high energy muons on transversely polarized proton and deuteron targets. 
Later, in Ref.~\cite{JeffersonLabHallA:2011ayy} the Jefferson Lab Hall A Collaboration
presented the first measurement of SSA in the SIDIS reaction $^3H(e,e',\pi^\pm,X)$ using 
transversely polarized target. 
For recent results of the COMPASS Collaboration, where Sivers asymmetries were extracted 
in SIDIS at the hard scales, see Refs.~\cite{COMPASS:2016led,COMPASS:2018ofp}. 
Since prediction of the Sivers asymmetry in Ref.~\cite{Brodsky:2002cx} 
big theoretical progress was achieved in further detailed understanding 
of this phenomena (see, e.g., Refs.~\cite{Brodsky:2002pr}-\cite{Ji:2020jeb}). 
A main advantage of the study of the Sivers asymmetry is its detailed access to the 
transverse momentum distribution function (TMD) and its spin-weighted moments.
As has been stressed in Ref.~\cite{Maji:2017wwd}, the model for the LFWFs 
was taken from Ref.~\cite{Gutsche:2013zia}. After Ref.~\cite{Gutsche:2013zia} we sufficiently 
improved the parametrization of the LFWFs to guarantee that the calculated parton distributions 
of the hadrons are consistent with model-independent scaling at large $x$. 
In particular, in Ref.~\cite{Gutsche:2016gcd}, which studies nucleon parton distributions 
and form factors, we showed that the parametrization of the LFWFs in terms of combinations of 
unpolarized and polarized PDFs produce the correct scaling for both the PDFs (by construction) 
as well as the TMDs and GPDs. The resulting nucleon form factors also obey the model-independent 
power scalings at large $Q^2$ discovered by 
Brodsky-Farrar-Matveev-Muradyan-Tavkhelidze~\cite{Brodsky:1973kr}. 
Later in series of papers (see Refs.~\cite{Sufian:2016hwn,deTeramond:2018ecg,%
Lyubovitskij:2020xqj,Lyubovitskij:2020otz}), a more accurate setup
for the LFWFs, consistent with model-independent constraints in QCD, were
successfully applied including gluon parton densities. 
Note that the model for nucleon LFWFs proposed in Ref.~\cite{Gutsche:2013zia} and used 
in Ref.~\cite{Maji:2017wwd} is not fully consistent with QCD constraints.  
That is why we have carried out the further improvements. We would also like to stress 
that the framework in Ref.~\cite{Maji:2017wwd} for the calculation of the asymmetries
in SIDIS is also not fully consistent. In particular, the authors of Ref.~\cite{Maji:2017wwd} 
do not calculate the fragmentation functions and use a phenomenological input, while these functions 
can be calculated directly in LF QCD as the other types of parton distribution. 
It is our main purpose that starting from the present paper, we intend to do a comprehensive analysis 
of the asymmetries in QCD processes at fixed targets, like Drell-Yan and SIDIS.

\vspace*{.25cm}

We start with basic notions of Holographic QCD approach based on quark-scalar diquark 
structure of nucleon. The $J^z = + \frac{1}{2}$ two-particle quark-scalar diquark Fock state 
is given by~\cite{Brodsky:2002cx,Brodsky:1980zm,Brodsky:2000ii,Gutsche:2016gcd,%
Gutsche:2013zia,Lyubovitskij:2020otz}: 
\eq 
\Big|\Psi^{\uparrow}(P^+,{\bf 0}_\perp)\Big\ra = 
\int \frac{d^2\bfk dx}{16 \pi^3 \sqrt{x (1-x)}} 
\, \biggl[ 
\varphi^{\uparrow}_{+\frac{1}{2}}(x,\bfk) \, \Big|+ \frac{1}{2}; x P^+, \bfk \Big\ra + 
\varphi^{\uparrow}_{-\frac{1}{2}}(x,\bfk) \, \Big|- \frac{1}{2}; x P^+, \bfk \Big\ra  
\biggr] 
\en 
where we use notation $\varphi^{\lambda_N}_{\lambda_u}(x,\bfk)$ for the light-front 
wave function (LFWF) at the initial scale $\mu_0$ with specific 
helicities for the proton $\lambda_N = \uparrow$ and $\downarrow$ and 
for the struck quark $\lambda_u = +\frac{1}{2}$ and 
$-\frac{1}{2}$~\cite{Gutsche:2016gcd,Gutsche:2013zia,Lyubovitskij:2020otz}: 
\eq
\varphi^{\uparrow}_{+ \frac{1}{2}}(x,\bfk) &=& \frac{2 \pi \sqrt{2}}{\kappa} \,
\sqrt{u_{v+}(x)}  \,
\exp\biggl[- \frac{\bfk^2}{2 \kappa^2}\biggr] \,,
\nonumber\\
\varphi^{\uparrow}_{- \frac{1}{2}}(x,\bfk) &=& 
- \frac{2 \pi \sqrt{2}}{\kappa^2} \, 
(k^1 + i k^2) \, 
\sqrt{u_{v-}(x)} \, 
\exp\biggl[- \frac{\bfk^2}{2 \kappa^2} \biggr] \,,
\en
where $u_{v\pm}(x) = u_v(x) \pm \delta u_v(x)$,
$u_v(x)$ and $\delta u_v(x)$ are the helicity-independent (unpolarized) 
and helicity-dependent (polarized) $u$ valence quark parton distributions, 
$\kappa \sim 350-500$ MeV is the scale dilaton parameter. 

In a similar way one can construct the $J^z = -1/2$ 
quark-scalar diquark Fock state components:
\eq\label{LFWF_Setup}
 \varphi^{\downarrow}_{+ \frac{1}{2}}(x,\bfk) &=& 
\frac{2 \pi \sqrt{2}}{\kappa^2} \, 
(k^1 - i k^2) \, 
\sqrt{u_{v-}(x)}  \,
\exp\biggl[- \frac{\bfk^2}{2 \kappa^2}\biggr] \,,
\nonumber\\
\varphi^{\downarrow}_{- \frac{1}{2}}(x,\bfk) &=& 
\frac{2 \pi \sqrt{2}}{\kappa} \,
\sqrt{u_{v+}(x)} \, 
\exp\biggl[- \frac{\bfk^2}{2 \kappa^2} \biggr] \,.
\en

The wave functions $\varphi^\uparrow_{\pm \frac{1}{2}}(x,\bfk)$
and $\varphi^\downarrow_{\pm \frac{1}{2}}(x,\bfk)$
are normalized in a such way that their combinations 
produce helicity-independent and helicity-dependent 
$u$-quark PDFs in the proton:
\eq
u_v(x) &=& 
\int \frac{d^2\bfk}{16\pi^3} \, 
\biggl[ |\varphi^{\uparrow}_{+ \frac{1}{2}}(x,\bfk)|^2 
      + |\varphi^{\uparrow}_{- \frac{1}{2}}(x,\bfk)|^2 
\biggr] \nonumber\\
&=& 
\int \frac{d^2\bfk}{16\pi^3} \, 
\biggl[ |\varphi^{\downarrow}_{+ \frac{1}{2}}(x,\bfk)|^2 
      + |\varphi^{\downarrow}_{- \frac{1}{2}}(x,\bfk)|^2 
\biggr] \,, \\ 
\delta u_v(x) &=& 
\int \frac{d^2\bfk}{16\pi^3} \, 
\biggl[ |\varphi^{\uparrow}_{+ \frac{1}{2}}(x,\bfk)|^2 
      - |\varphi^{\uparrow}_{- \frac{1}{2}}(x,\bfk)|^2 
\biggr] \nonumber\\
&=& 
\int \frac{d^2\bfk}{16\pi^3} \, 
\biggl[ |\varphi^{\downarrow}_{- \frac{1}{2}}(x,\bfk)|^2 
      - |\varphi^{\downarrow}_{+ \frac{1}{2}}(x,\bfk)|^2 
\biggr] 
\,.
\en
 
Note that in the original version of the quark-scalar diquark 
model~\cite{Brodsky:2002cx,Brodsky:1980zm,Brodsky:2000ii} by construction 
the proton LFWFs $|u[ud]_0\ra$ has $L=0$ and $L=1$ components with equal probability. 
In Ref.~\cite{Gutsche:2016gcd} it was proposed to relate them with two 
combinations of valence quark PDFs. It led to not equal probabilities 
of the $L=0$ and $L=1$ components. 

Next we define four possible spin-spin transitions between proton and struck quark: 
\eq 
\Uparrow \Big(J_p^z= + \frac{1}{2}\Big) &\to& \uparrow \Big(J_q^z = + \frac{1}{2}\Big) 
\,, \nonumber\\
\Downarrow \Big(J_p^z= - \frac{1}{2}\Big) &\to& \uparrow \Big(J_q^z = + \frac{1}{2}\Big) 
\,, \nonumber\\
\Uparrow \Big(J_p^z= + \frac{1}{2}\Big) &\to& \downarrow \Big(J_q^z = - \frac{1}{2}\Big)  
\,, \nonumber\\
\Downarrow \Big(J_p^z= - \frac{1}{2}\Big) &\to& \downarrow \Big(J_q^z = - \frac{1}{2}\Big) 
\,.
\en 
The corresponding amplitudes ${\cal A}$ including both tree and one-loop contributions 
have been calculated in Ref.~\cite{Brodsky:2002cx} 
using diquark model. Here, we present our results for the ${\cal A}$'s using Holographic QCD: 
\eq 
& &{\cal A}(\Uparrow \to \uparrow) = 
C^+(\Delta) \, h(\bfl^2)  \ e^{i \chi_1}  
\,, \nonumber\\[-.5mm]
& &{\cal A}(\Downarrow \to \uparrow) = \frac{\ell^1 - i \ell^2}{\kappa} \, 
C^-(\Delta) \, h(\bfl^2)  \ e^{i \chi_2}  
\,, \nonumber\\[-.5mm]
& &{\cal A}(\Uparrow \to \downarrow) = - \frac{\ell^1 + i \ell^2}{\kappa}
C^-(\Delta) \, h(\bfl^2)  \ e^{i \chi_2}  
\,, \nonumber\\[-.5mm]
& &{\cal A}(\Downarrow \to \downarrow) = 
C^+(\Delta) \, h(\bfl^2)  \ e^{i \chi_1}  
\,. 
\en 
Here  
\eq\label{Cpm} 
C^\pm(\Delta)   &=& 2 P^+ \, g \, e_u \, 
\Big(u_{v\pm}(\Delta) \, \Delta (1-\Delta)\Big)^{1/2} \,, 
\nonumber\\
h(\bfl^2) &=& \frac{2 \pi \sqrt{2}}{\kappa} \,
\exp\biggl[- \frac{\bfl^2}{2 \kappa^2} \biggr] \,.
\en 
and $\chi_{1,2}$ are the rescattering phases taking into account one-loop 
contributions~\cite{Brodsky:2002cx}: 
\eq 
\chi_i = \frac{\alpha_s C_F }{2} \, g_i \,, 
\en 
$g$ is the coupling of proton and $u[ud]_0$ quark-scalar diquark state, 
$\Delta = k^+/P^+$, $\bfl = (\ell^x,\ell^y)$, and 
$\alpha_s = \alpha_s(\mu^2) = 4 \pi \beta_0/\log(\mu^2/\Lambda_{\rm QCD}^2)$ is 
the QCD coupling at the scale $\mu$, $C_F = 4/3$, $\beta_0 = 9$, 
$\Lambda_{\rm QCD} = 0.226$ GeV.  
As shown in Eq.~(\ref{LFWF_Setup}) we have parametrized the LFWFs in terms of combinations
of quark PDFs $u_{v\pm}$, instead of the simple ansatz of the original spectator model.
Therefore, in Eq.~(\ref{LFWF_Setup}) the combination $u_{v+}$ appears in the case of 
the spin-spin transitions without flip, while $u_{v-}$ appears in case of the spin-spin 
transitions with spin flip. Therefore, the $C^+$ is proportional to $u_{v+}$ and 
$C^-$ to $u_{v-}$.

The loop couplings $g_1$ and $g_2$ have been calculated in 
in Ref.~\cite{Brodsky:2002cx}. It was shown that, while $g_1$ and $g_2$ depend in the 
infrared regulator $\lambda_g$ (gluon mass), their difference is independent of $\lambda_g$ 
and, therefore, is infrared finite: 
\eq 
\Delta g &=& g_1 - g_2 = \frac{\kappa^2}{\bfl^2} \, 
\log\frac{\bfl^2 + B(\Delta)}{B(\Delta)} \,, \nonumber\\
B(\Delta) &=& m^2 (1-\Delta) + m_{[ud]}^2 \Delta - M_N^2 \Delta (1-\Delta) \,, 
\en 
where $m$ and $m_{[ud]}$ are the constituent masses of struck $u$ quark and 
spectator diquark $[ud]_0$, $M_N$ is the nucleon mass.  

Note, the $\Delta$ contributes to 
the  azimuthal single-spin asymmetry transverse to the production 
plane~\cite{Brodsky:2002cx}: 
\eq 
   {\cal P}_y = \alpha_s \, C_F \, 
\dfrac{\ell^x}{\kappa} \, \dfrac{\Delta g \,  \sqrt{1 - R^2(\Delta)}}{1+R(\Delta)
+\dfrac{\bfl^2}{\kappa^2} (1-R(\Delta))} \,,
\en 
where $R(\Delta)=\delta u_v(\Delta)/u_v(\Delta)$. 
The novel result is that we expressed the ${\cal P}_y$ asymmetry through the ratio of the 
valence PDFs of the $u$ quark --- $\delta u_v$ and $u_v$.  

Next, following Ref.~\cite{Hwang:2010dd} we can calculate the Sivers $f_{1T}^\perp(x,\bfk)$ 
and Boer-Mulders $h_1^\perp(x,\bfk)$ 
T-odd $u$-quark TMDs. First, we update the LFWFs including rescattering 
phases $\chi_{1,2}$ due to final state interactions~\cite{Hwang:2010dd}:  
\eq\label{Psi_LFWFs}
\Phi^{\uparrow}_{+ \frac{1}{2}}(x,\bfk) &=& \varphi^{\uparrow}_{+ \frac{1}{2}}(x,\bfk) \, 
\, e^{i \chi_1} \,, \nonumber\\
\Phi^{\uparrow}_{- \frac{1}{2}}(x,\bfk) &=& \varphi^{\uparrow}_{- \frac{1}{2}}(x,\bfk) \, 
\, e^{i \chi_2} \,, \nonumber\\
\Phi^{\downarrow}_{+ \frac{1}{2}}(x,\bfk) &=& \varphi^{\downarrow}_{+ \frac{1}{2}}(x,\bfk) \, 
\, e^{i \chi_2} \,, \nonumber\\
\Phi^{\downarrow}_{- \frac{1}{2}}(x,\bfk) &=& \varphi^{\downarrow}_{- \frac{1}{2}}(x,\bfk) \, 
\, e^{i \chi_1} \,.
\en 
Second, we use the light-front decomposition for the Sivers and Boer-Mulders 
functions in terms of the LFWFs from Eq.~(\ref{Psi_LFWFs}): 
\eq 
\frac{k^x}{M_N} \, f_{1T}^{\perp u}(x,\bfk) &\equiv& 
- \frac{i}{32 \pi^3} \, \biggl[ 
    \Phi^{\dagger \uparrow}_{+ \frac{1}{2}}(x,\bfk) \Phi^{\downarrow}_{+ \frac{1}{2}}(x,\bfk) 
  + \Phi^{\dagger \uparrow}_{- \frac{1}{2}}(x,\bfk) \Phi^{\downarrow}_{- \frac{1}{2}}(x,\bfk) 
\nonumber\\
  &-& \Phi^{\dagger \downarrow}_{- \frac{1}{2}}(x,\bfk) \Phi^{\uparrow}_{- \frac{1}{2}}(x,\bfk) 
  - \Phi^{\dagger \downarrow}_{+ \frac{1}{2}}(x,\bfk) \Phi^{\uparrow}_{+ \frac{1}{2}}(x,\bfk) 
\biggr] \,, \\
\frac{k^x}{M_N} \, h_1^{\perp u}(x,\bfk) &=&  
- \frac{i}{32 \pi^3} \, \biggl[ 
    \Phi^{\dagger \uparrow}_{- \frac{1}{2}}(x,\bfk) \Phi^{\uparrow}_{+ \frac{1}{2}}(x,\bfk) 
  - \Phi^{\dagger \uparrow}_{+ \frac{1}{2}}(x,\bfk) \Phi^{\uparrow}_{- \frac{1}{2}}(x,\bfk) 
\nonumber\\
&+& \Phi^{\dagger \downarrow}_{- \frac{1}{2}}(x,\bfk) \Phi^{\downarrow}_{+ \frac{1}{2}}(x,\bfk) 
 -  \Phi^{\dagger \downarrow}_{+ \frac{1}{2}}(x,\bfk) \Phi^{\downarrow}_{- \frac{1}{2}}(x,\bfk) 
\biggr] \,. 
\en  
We get 
\eq 
f_{1T}^{\perp u}(x,\bfk) \equiv  h_1^{\perp u}(x,\bfk) 
= - \frac{\alpha_s \, C_F}{2 \pi \kappa^3} \, M_N  \, \Delta g 
\, \sqrt{u_v^2(x) - \delta u_v^2(x)}
\, \exp\Big[ - \frac{\bfk^2}{\kappa^2}\Big] \,. 
\en 
Using our result for the unpolarized TMD $f_1(x,\bfk)$~\cite{Gutsche:2016gcd}: 
\eq 
f_1^u(x,\bfk) &=& \frac{1}{32 \pi^3} \, \biggl[ 
    |\Phi^{\uparrow}_{+ \frac{1}{2}}(x,\bfk)|^2 
  + |\Phi^{\downarrow}_{+ \frac{1}{2}}(x,\bfk)|^2 
  + |\Phi^{\uparrow}_{- \frac{1}{2}}(x,\bfk)|^2 
  + |\Phi^{\downarrow}_{- \frac{1}{2}}(x,\bfk)|^2 
\biggr] \nonumber\\
&=& \frac{1}{2 \pi \kappa^2} \, 
\biggl[ u_v(x) + \delta u_v(x) + \dfrac{\bfk^2}{\kappa^2} \, 
\Big(u_v(x) - \delta u_v(x)\Big) \biggr] \, \exp\Big[ - \frac{\bfk^2}{\kappa^2}\Big] 
\en 
we reproduce the well-known relation between the asymmetry ${\cal P}_y$ and 
TMDs $f_1^u(x,\bfk)$ and $f_{1T}^{\perp u}(x,\bfk)$~\cite{Brodsky:2002rv,Boer:2002ju}: 
\eq 
{\cal P}_y = - \frac{\ell^x}{M_N} \, \frac{f_{1T}^{\perp u}(\Delta,\bfl)}{f_1^u(\Delta,\bfl)} \,. 
\en  
Notice that the results for the $d$-quark TMDs are obtained upon replacement
$u_v(x) \to d_v(x)$ and $\delta u_v(x) \to \delta d_v(x)$ and changing overall sign in 
$\varphi^{\uparrow}_{- \frac{1}{2}}(x,\bfk)$ and 
$\varphi^{\downarrow}_{+ \frac{1}{2}}(x,\bfk)$~\cite{Gutsche:2016gcd}.
One gets: 
\eq
f_1^d(x,\bfk) &=& \frac{1}{2 \pi \kappa^2} \, 
\biggl[ d_v(x) + \delta d_v(x) + \dfrac{\bfk^2}{\kappa^2} \,
\Big(d_v(x) - \delta d_v(x)\Big) \biggr] \, \exp\Big[ - \frac{\bfk^2}{\kappa^2}\Big]\,, 
\nonumber\\
f_{1T}^{\perp d}(x,\bfk) &\equiv&  h_1^{\perp d}(x,\bfk)
= \frac{\alpha_s \, C_F}{2 \pi \kappa^3} \, M_N  \, \Delta g
\, \sqrt{d_v^2(x) - \delta d_v^2(x)}
\, \exp\Big[ - \frac{\bfk^2}{\kappa^2}\Big] \,.
\en
Note, that in the scalar quark-diquark model, the relations between the Sivers and 
Boer-Mulders functions for both flavors ($u$ and $d$ quarks) have the same 
sign~\cite{Bacchetta:2008af,Hwang:2010dd}. 
It is true that inclusion of the axial-vector diquark will distinguish
the relations for $u$ and $d$ quark. See Refs.~~\cite{Bacchetta:2008af,Hwang:2010dd}. 

Now we present our numerical results for the Sivers asymmetry 
and TMDs. In our analysis, 
we will use the following set of parameters: $\mu_0^2 = 0.40$ GeV$^2$ is the 
initial scale, 
$M_N = 0.93827$ GeV is the nucleon mass, $m=0.35$ GeV is the constituent mass 
of $u$ and $d$ quark, and $\kappa = 0.5$ GeV is the dilaton scale parameter. 
The mass of the $[ud]_0$ scalar diquark $m_{[ud]}$ will be varied from 
0.65 to 0.8 GeV. For the $q_v$ and $\delta q_v$ ($q=u,d$) quark PDFs and the  
ratio $R(x) = \delta u_v(x)/u_v(x)$ we use the results of the 
Gl\"uck-Reya-Vogt (GRV)/Gl\"uck-Reya-Stratmann-Vogelsang (GRSV) fit done 
in Refs.~\cite{Gluck:1998xa,Gluck:2000dy}. In particular,  
$R(x) \simeq 2.043 \, x^{0.97} \, (1-x)^{0.64}$. 

\begin{table}[ht]
\begin{center}
\caption{First moments of the Sivers TMDs for $u$ and $d$ quark} 
\vspace*{.1cm}

\def\arraystretch{1.25}
    \begin{tabular}{|c|c|c|c|c|c|}
      \hline
$x$ & $Q^2$ (GeV$^2$) & \multicolumn{2}{c|}{$|x f_{1T}^{\perp (1)u}(x)|$}  
                      & \multicolumn{2}{c|}{$|x f_{1T}^{\perp (1)d}(x)|$} \\ 
\cline{3-6}
    & & COMPASS~\cite{COMPASS:2018ofp} & Our results
      & COMPASS~\cite{COMPASS:2018ofp} & Our results \\ 
\hline
0.0063 & 1.27 & $|0.0022 \pm 0.0051|$ 
              & $0.0004  \pm 0.0001$ 
              & $|0.001  \pm 0.021|$        
              & $0.0002  \pm 0.00001$ \\
0.0105 & 1.55 & $|0.0029 \pm 0.0040|$ 
              & $0.0007  \pm 0.0001$ 
              & $|0.004  \pm 0.017|$        
              & $0.0003  \pm 0.0001$ \\
0.0164 & 1.83 & $0.0058  \pm 0.0037$ 
              & $0.0013  \pm 0.0002$ 
              & $0.019   \pm 0.015$        
              & $0.0005  \pm 0.0001$ \\
0.0257 & 2.17 & $0.0097  \pm 0.0033$ 
              & $0.0022  \pm 0.0003$ 
              & $0.034   \pm 0.013$        
              & $0.0008  \pm 0.0002$ \\
0.0399 & 2.82 & $0.0179  \pm 0.0036$ 
              & $0.004   \pm 0.005$ 
              & $0.032   \pm 0.015$        
              & $0.0014  \pm 0.0003$ \\
0.0629 & 4.34 & $0.0224  \pm 0.0046$ 
              & $0.007   \pm 0.001$ 
              & $0.048   \pm 0.019$        
              & $0.0025  \pm 0.0005$ \\
0.101  & 6.76 & $0.0171  \pm 0.0057$ 
              & $0.012   \pm 0.003$ 
              & $0.025   \pm 0.023$        
              & $0.004   \pm 0.001$ \\
0.163  & 10.6 & $0.0295  \pm 0.0070$ 
              & $0.018   \pm 0.004$ 
              & $0.056   \pm 0.027$        
              & $0.006   \pm 0.002$ \\
0.288  & 20.7 & $0.0160  \pm 0.0073$ 
              & $0.019   \pm 0.006$ 
              & $|0.017   \pm 0.028|$        
              & $0.005   \pm 0.002$ \\
\hline
\end{tabular}
\label{Tab:1}
\end{center}
\end{table}

\begin{figure} 
\begin{center}
\epsfig{figure=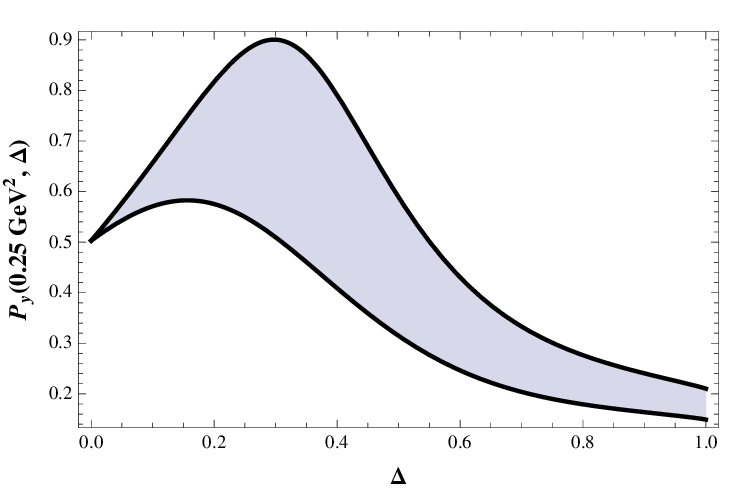,scale=.5}
\hspace*{1cm}
\epsfig{figure=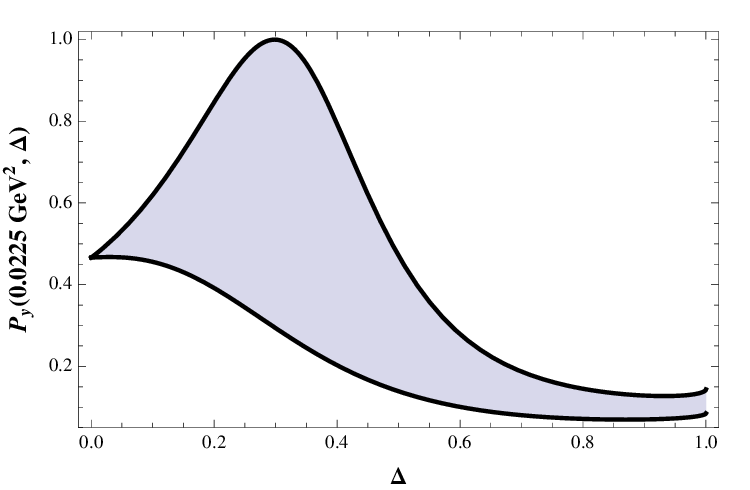,scale=.5}
\end{center}
\vspace*{-.48cm}
\noindent
\caption{2D plots of the ${\cal P}_y$ asymmetry as function of 
$\Delta$ at $|\bfl| = 0.5$ GeV (left panel) 
and $|\bfl| = 0.15$ GeV (right panel). 
\label{fig3}}

\begin{center}
\epsfig{figure=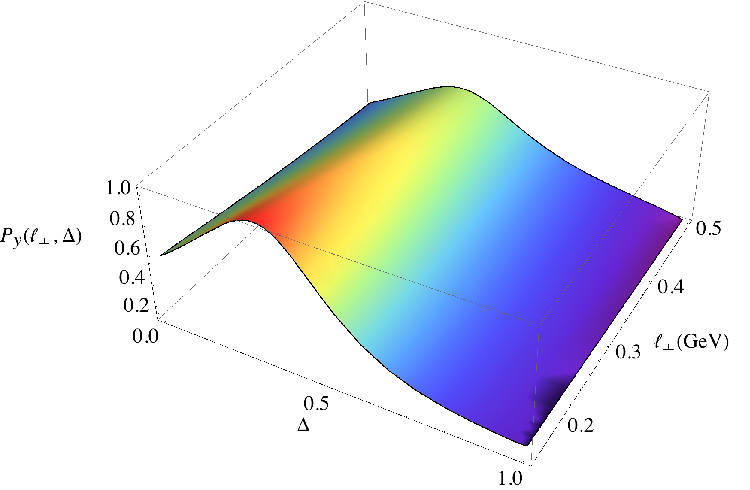,scale=.57}
\hspace*{1cm}
\epsfig{figure=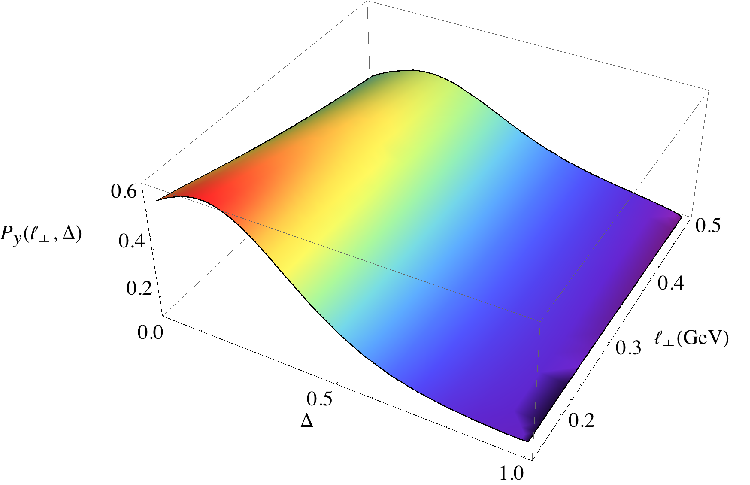,scale=.57}
\end{center}
\vspace*{-.48cm}
\noindent
\caption{3D plots of the ${\cal P}_y$ asymmetry as function of 
$\Delta$ and $|\bfl|$ for $m_{[ud]} = 0.65$ GeV (left panel)  
and for $m_{[ud]} = 0.8$ GeV (right panel). 
\label{fig4}}

\begin{center}
\epsfig{figure=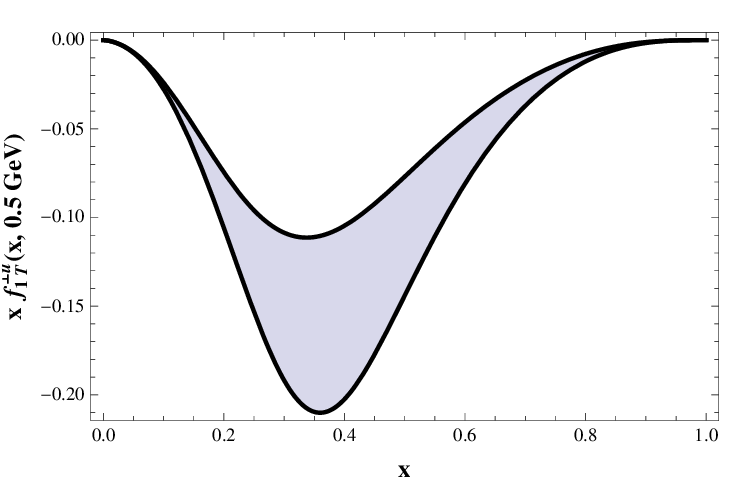,scale=.5}
\hspace*{1cm}
\epsfig{figure=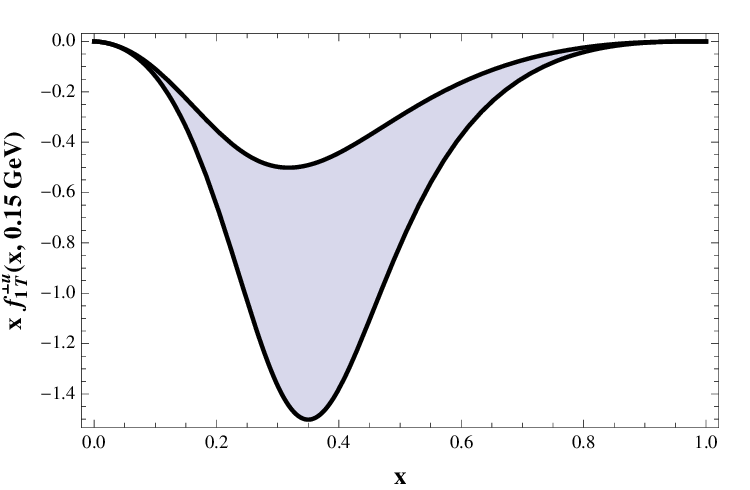,scale=.5}
\end{center}
\vspace*{-.48cm}
\noindent
\caption{2D plots of the Sivers TMD $x f_{1T}^{\perp u}(x,\bfk)$ 
as function of
$x$ at $|\bfk| = 0.5$ GeV (left panel)
and $|\bfk| = 0.15$ GeV (right panel).
\label{fig5}}

\begin{center}
\epsfig{figure=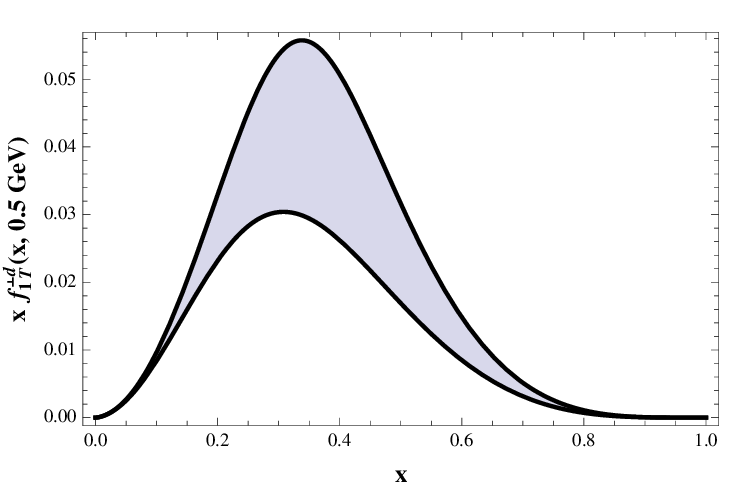,scale=.5}
\hspace*{1cm}
\epsfig{figure=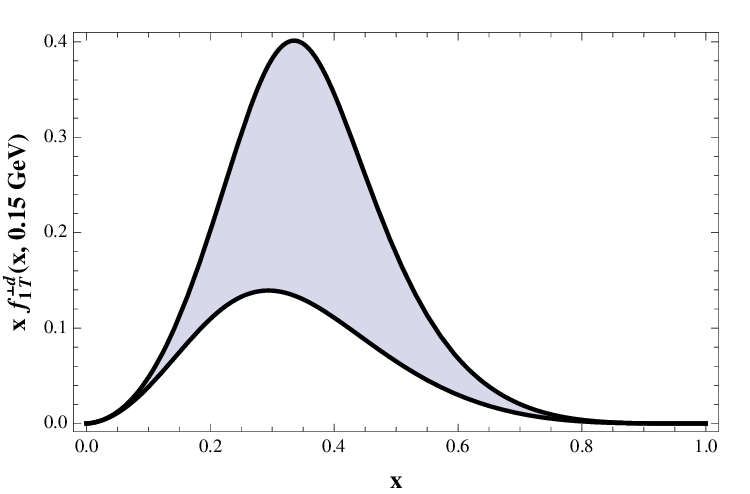,scale=.5}
\end{center}
\vspace*{-.48cm}
\noindent
\caption{2D plots of the Sivers TMD 
$x f_{1T}^{\perp d}(x,\bfk)$ as function of
$x$ at $|\bfk| = 0.5$ GeV (left panel)
and $|\bfk| = 0.15$ GeV (right panel).
\label{fig6}}
\end{figure}

In Fig.~\ref{fig3} we show two-dimensional plots for the ${\cal P}_y$ asymmetry 
as function of $\Delta$ variable at $|\bfl| = 0.5$ GeV $|\bfl| = 0.15$ GeV. 
The shaded bands correspond to a variation of the $m_{[ud]} = 0.65-0.8$ GeV. 
Increasing of $m_{[ud]}$ leads to a suppression of the asymmetry ${\cal P}_y$. 
In Fig.~\ref{fig4} we present 
three-dimensional plots for the ${\cal P}_y$ as function of two variables 
$\Delta$ and $|\bfl|$ running from 0 to 1 and to 0.15 to 0.5 GeV, respectively, 
at two typical (low and upper) values of the scalar diquark mass $m_{[ud]} = 0.65$ 
and 0.8 GeV. 

\begin{figure}
\begin{center}
\epsfig{figure=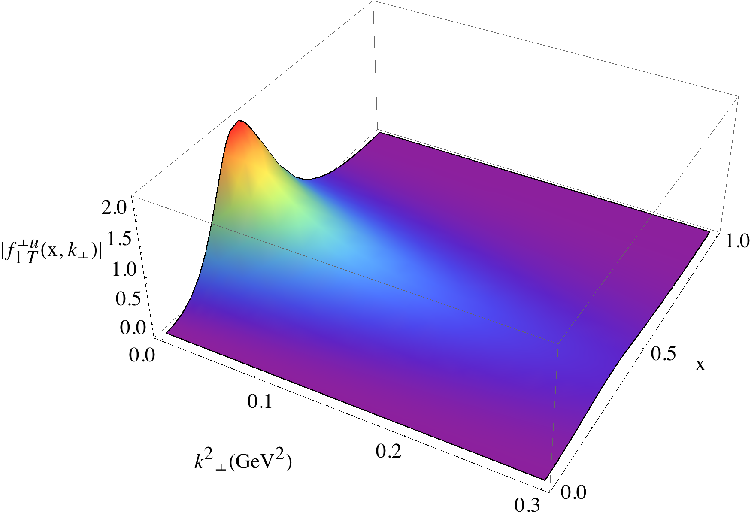,scale=.57}
\hspace*{1cm}
\epsfig{figure=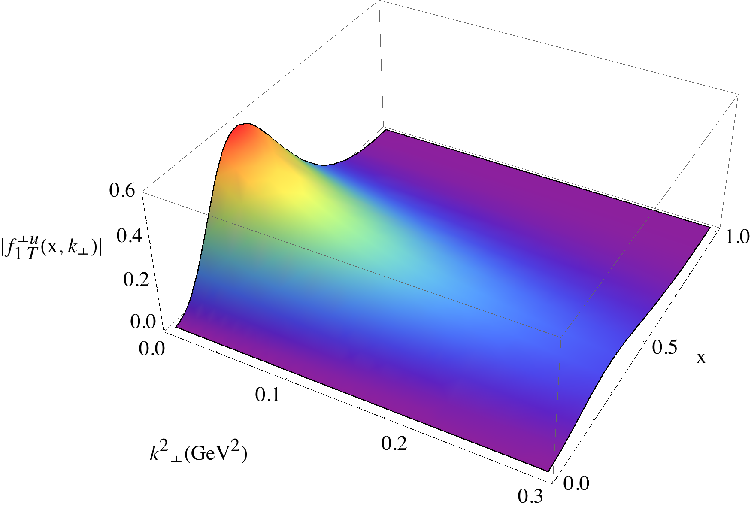,scale=.57}
\end{center}
\vspace*{-.48cm}
\noindent
\caption{3D plots of the Sivers TMD $|x f_{1T}^{\perp u}(x,\bfk)|$ 
as function of $x$ and $\bfk^2$ for $m_{[ud]} = 0.65$ GeV (left panel)  
and for $m_{[ud]} = 0.8$ GeV (right panel). 
\label{fig7}}

\begin{center}
\epsfig{figure=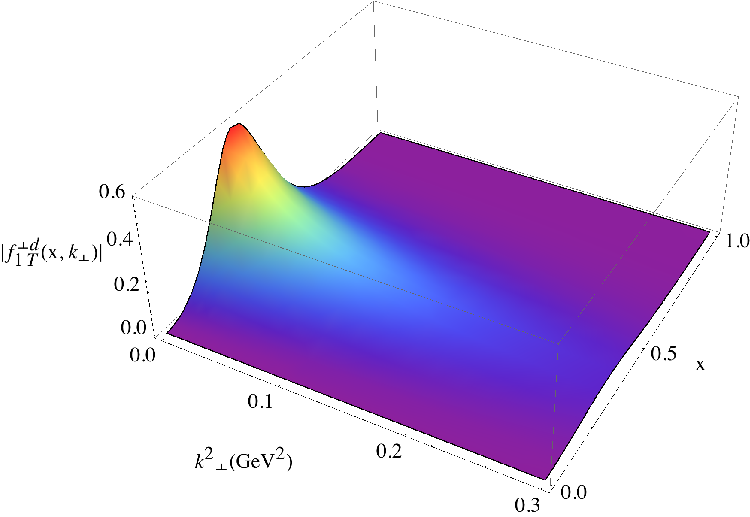,scale=.57}
\hspace*{1cm}
\epsfig{figure=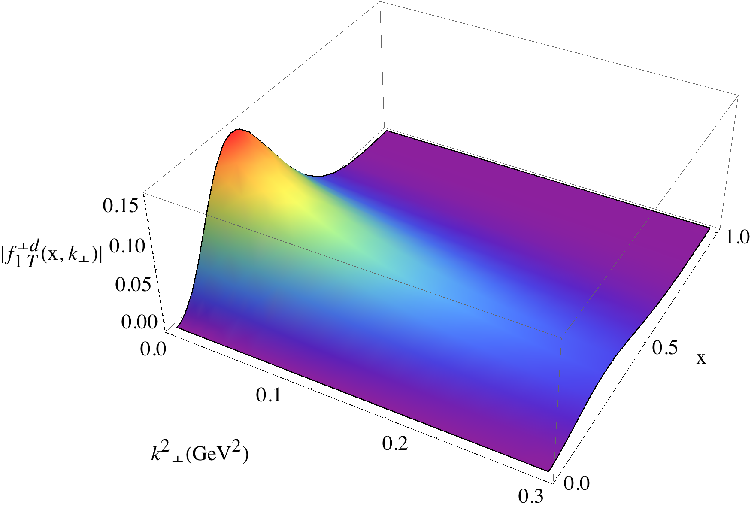,scale=.57}
\end{center}
\vspace*{-.48cm}
\noindent
\caption{3D plots of the Sivers TMD $|x f_{1T}^{\perp d}(x,\bfk)|$ 
as function of $x$ and $\bfk^2$ for $m_{[ud]} = 0.65$ GeV (left panel)  
and for $m_{[ud]} = 0.8$ GeV (right panel). 
\label{fig8}}

\begin{center}
\epsfig{figure=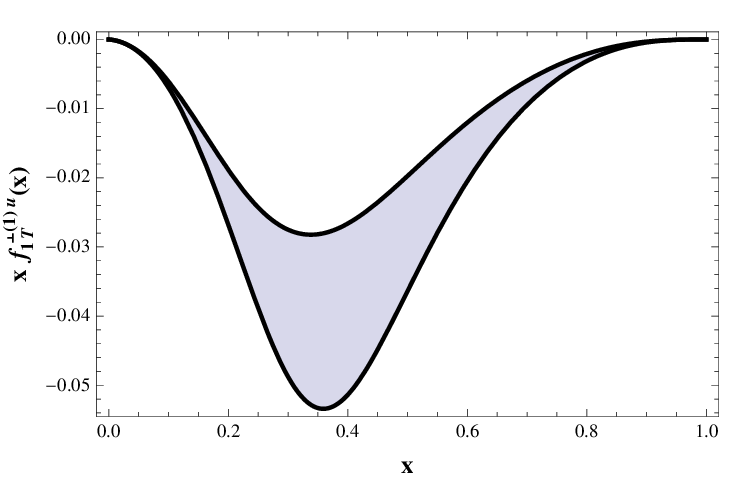,scale=.5}
\hspace*{1cm}
\epsfig{figure=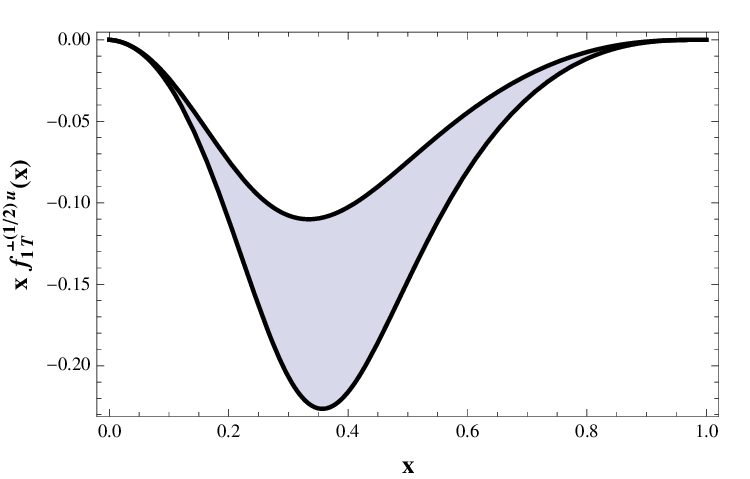,scale=.5}
\end{center}
\vspace*{-.48cm}
\noindent
\caption{2D plots of the moments of the Sivers TMD 
$x f_{1T}^{\perp (1)u}(x)$ and $x f_{1T}^{\perp (1/2)u}(x)$. 
\label{fig9}}

\begin{center}
\epsfig{figure=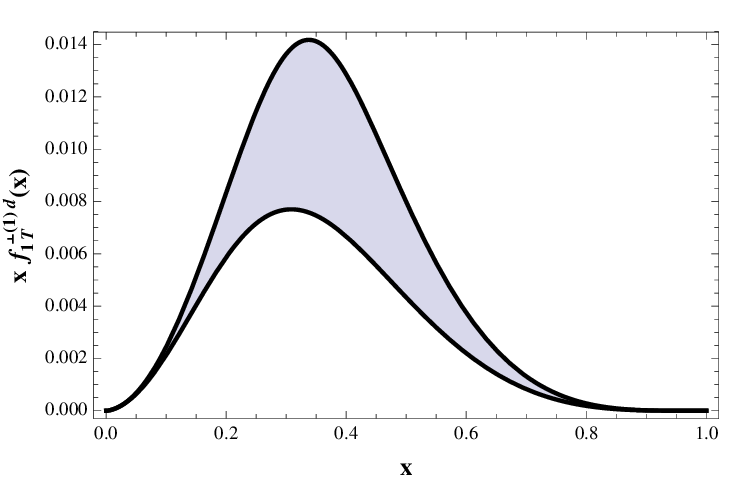,scale=.5}
\hspace*{1cm}
\epsfig{figure=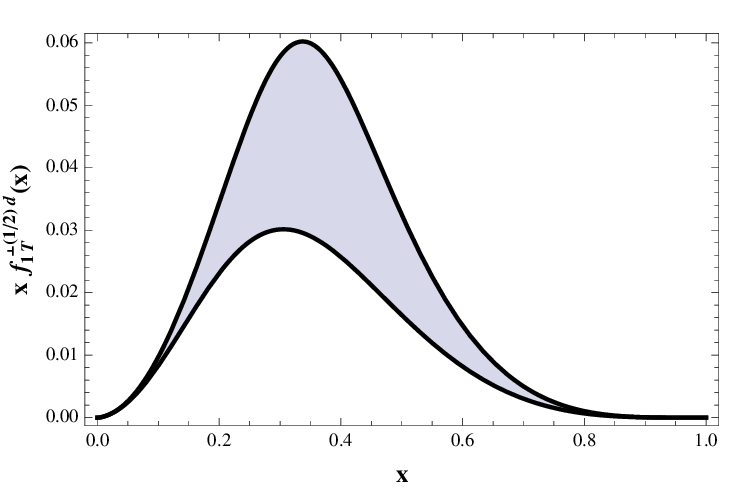,scale=.5}
\end{center}
\vspace*{-.48cm}
\noindent
\caption{2D plots of the moments of the Sivers TMD 
$x f_{1T}^{\perp (1)d}(x)$ and $x f_{1T}^{\perp (1/2)d}(x)$. 
\label{fig10}}
\end{figure}

In Figs.~\ref{fig5}-\ref{fig10} we present our results for the 
$u$ and $d$ quark Sivers TMDs and for their first and $1/2$ moments: 
\eq
f_{1T}^{\perp (1)q}(x) &=& \int d^2\bfk \, \frac{\bfk^2}{2 M_N^2} \, f_{1T}^{\perp q}(x,\bfk) 
\nonumber\\
&=& - \alpha_s \, C_F \, \frac{\kappa}{4 M_N} \, n_q 
\, \sqrt{q_v^2(x) - \delta q_v^2(x)} 
\,\, \Gamma[0,b(x)]  
\, \exp[b(x)] 
\,, \nonumber\\
f_{1T}^{\perp (1/2)q}(x) &=& \int d^2\bfk \, \frac{|\bfk|}{M_N} \, f_{1T}^{\perp q}(x,\bfk) 
\nonumber\\
&=& - \alpha_s \, C_F \, n_q
\, \sqrt{q_v^2(x) - \delta q_v^2(x)} \ \int\limits_0^\infty dt 
\, \exp[-t^2] \, \log\dfrac{t^2+b(x)}{b(x)} 
\,, 
\en 
where $n_u = 1$, $n_d = -1$, $b(x)=B(x)/\kappa^2$, 
\eq 
\Gamma[a,z] = \int\limits_z^\infty dt \, t^{a-1} \, e^{-t}
\en 
is the incomplete Gamma function. 
In particular, in Figs.~\ref{fig5} and~\ref{fig6} 
we show our predictions for the Sivers TMD $x f_{1T}^{\perp u}(x,\bfk)$ 
and $x f_{1T}^{\perp d}(x,\bfk)$ as functions of
$x$ at $|\bfk| = 0.5$ and 0.15 GeV. In Figs.~\ref{fig7} and~\ref{fig8} 
we display 
three-dimensional plots for the Sivers TMDs 
$|f_{1T}^{\perp u}(x,\bfk)|$ and $|f_{1T}^{\perp d}(x,\bfk)|$  
as a functions of $x$ and $\bfk^2$ for $m_{[ud]} = 0.65$ and $0.8$ GeV. 
Finally, in Figs.~\ref{fig9} and~\ref{fig10} we present our results for 
the first and $1/2$ moments of the Sivers TMDs. 

In Table~\ref{Tab:1} we compare 
our results for the magnitudes of the first moments of the Sivers TMDs for $u$ 
and $d$ quarks $|x f_{1T}^{\perp (1)u,d}(x)|$ with data 
extracted by the COMPASS Collaboration~\cite{COMPASS:2018ofp} at different scales 
of resolution $Q^2$. Note, that our results for the moments 
of the Sivers quark TMDs are in good agreement with results of  
other theoretical approaches (see, e.g, compilation of theoretical 
predictions in Ref.~\cite{Anselmino:2005an}). 

In conclusion, we have presented predictions for the Sivers asymmetry and TMDs 
based on holographic QCD. These results will provide an important test of 
this novel approach to nonperturbative hadron dynamics. 

\begin{acknowledgments}

This work was funded by BMBF (Germany) ``Verbundprojekt 05P2021 (ErUM-FSP T01) -
Run 3 von ALICE am LHC: Perturbative Berechnungen von Wirkungsquerschnitten
f\"ur ALICE'' (F\"orderkennzeichen: 05P21VTCAA), by ANID PIA/APOYO AFB180002 (Chile),
by FONDECYT (Chile) under Grant No. 1191103
and by ANID$-$Millen\-nium Program$-$ICN2019\_044 (Chile).
The work of S.J.B. was supported in part by the Department of Energy 
under contract DE-AC02-76SF00515. SLAC-PUB-17676.

\end{acknowledgments}

\clearpage 

\clearpage 

\end{document}